\documentclass[aps,prb,twocolumn,superscriptaddress]{revtex4}

\usepackage{amsmath}
\usepackage{amssymb}
\usepackage{wasysym}
\usepackage{psfrag}
\usepackage{tabularx}
\usepackage[dvips]{graphicx}
\usepackage{enumitem}
\usepackage{color}
\def\beq{\begin{equation}}
\def\eeq{\end{equation}}
\def\bea{\begin{eqnarray}}
\def\eea{\end{eqnarray}}

\newcommand{\Sp}{\mathbf{S}}
\newcommand{\Tp}{\mathbf{s}}
\def\rv {{\bf r}}

\def\Vop {\hat{{\bf v}}}
\def\Rop {\hat{{\bf R}}}

\begin{document}

\title{From frustrated magnetism to spontaneous  Chern insulators}
\author{H. D. Rosales}
\email[corresponding author: ]{rosales@fisica.unlp.edu.ar}
\affiliation{Instituto de F\'isica de L\'iquidos y Sistemas Biol\'ogicos (IFLYSIB), UNLP-CONICET, La Plata, Argentina and Departamento de F\'isica, Facultad de Ciencias Exactas,
Universidad Nacional de La Plata, c.c. 16, suc. 4, 1900 La Plata, Argentina.}
\affiliation{Facultad de Ingenier\'ia, UNLP, La Plata, Argentina}
\author{F.A. G\'omez Albarrac\'in}
\email[]{albarrac@fisica.unlp.edu.ar}
\affiliation{Instituto de F\'isica de L\'iquidos y Sistemas Biol\'ogicos (IFLYSIB), UNLP-CONICET, La Plata, Argentina and Departamento de F\'isica, Facultad de Ciencias Exactas,
Universidad Nacional de La Plata, c.c. 16, suc. 4, 1900 La Plata, Argentina.}
\author{P. Pujol}
\email[]{pierre.pujol@irsamc.ups-tlse.fr}
\affiliation{Laboratoire de Physique Theorique-IRSAMC, CNRS and Universit\'e de Toulouse, UPS, Toulouse, F-31062, France.}

\begin{abstract}
We study the behaviour of electrons interacting with a classical magnetic background via a strong Hund coupling. 
The magnetic background results from a Hamiltonian which favours at low temperature the emergence of a phase with non-zero scalar chirality.
The strong Hund's coupling combined with the total chirality of the classical spins induces in the electrons an effective flux which results
in the realisation of a band structure with non zero Chern number.
First, we consider as a magnetic background a classical spin system with spontaneous net chirality.
We study the Density of States (DoS) and Hall conductance  in order to analyse the topological transitions, as a function of the Fermi energy and the temperature
of the classical spins. We also study a similar model in which the Chern number of the filled bands can be ``tuned'' with the external magnetic field, resulting in a topological insulator 
in which the direction of the chiral edge mode can be reverted by reversing the orientantion of the magnetic field applied to the classical magnetic system. 
\end{abstract}

\maketitle

\section{Introduction}
The discovery of the quantum Hall effect in 1980 \cite{Quantum_Hall} opened a new field in condensed matter physics. 
In this area, topological phases have attracted much interest in the last decade and in particular, the celebrated topological insulator
(for a review on the subject see [\onlinecite{Topo_Ins}]). 
The first topological insulator correspond to the integer quantum Hall 
effect  where a bidimensional electron gas in a  strong magnetic field $B$ has a Hall conductance with quantised plateaux at
$\sigma_{xy}=\nu e^2/h$ ($\nu=1,2,...$) values.

The quantum anomalous Hall (QAH) effect is a phenomenon in solids arising from the spin-orbit coupling due to an interaction between 
the electronic spin and local magnetic moments. It can lead to a topologically non-trivial electronic structure with a quantized Hall effect 
characterised by the Chern numbers $C$ of the filled bands and the presence of chiral edge modes. The principal ingredient is a non-collinear spin texture
distinguished by a non zero value of the uniform scalar chirality parameter \cite{Nagaosa_1,Taillefumier,Nagaosa_2}.  

Experimentally, the QAH effect was first observed in thin films of chromium-doped (Bi,Sb)$_2$Te$_3$,
a material known to be a magnetic topological insulator \cite{Science}. Much work has been done since then. For recent reviews on experimental advances,
see [\onlinecite{Review1,Review2}].

In this direction, frustrated systems are good candidates for the emergence of chiral spin textures. Among these, kagome materials \cite{KagomeMaterials} 
have been studied both theoretically and experimentally \cite{BookLacroix} due to the possibility for nontrivial spin textures, spin-liquid phases 
\cite{ShengBecca}, fractional Chern insulators  \cite{Fradkin}, exotic  transport properties, and topologically protected phases (for a recent review  see Ref.[\onlinecite{Norman2016}]
and references therein). Furthermore, quite recently the QAH effect has been measured in the material Co$_3$Sn$_2$S, where the mangetic structure
consists of stacked kagome lattices. This material is a Weyl semimetal and is proposed to be a strong candidate to observe the QAH state in a two dimensional system,
as stated in Ref.[\onlinecite{NatPhys18}].

Regarding the possible emergence of chiral phases in the kagome lattice, in a recent paper \cite{Flavia+Pierre} 
the authors found that a spontanesouly broken reflexion symmetry phase with a non zero total
chirality can be induced by a magnetic field. They found 
that by decreasing the temperature the system undergoes a phase transition from a normal paramagnet to a chiral state, 
with either positive or negative total chirality.  
This opens the possibility of triggering the anomalous Hall effect and the emergence of a Chern insulators by just cooling and
applying a magnetic field to the magnetic system.

Motivated by the previous discussion, in this paper we focus on a model described by the Kondo
lattice model on the kagome lattice combining electronic and magnetic degrees of freedom. 
Its first ingredient is given by an electronic term consisting in non-interacting electrons 
evolving in a classical magnetic background on the kagome lattice. There is a Hund's coupling between the electronic spin and the local moments 
on each site of the lattice. The second ingredient governs the precise shape of the classical magnetic background, 
is given by a  pure spin Hamiltonian.  We will focus in two models: an XXZ up to 
second nearest neighbours and a Heisenberg model with Dzyaloshinskii-Moriya interaction. 
In the first case, by cooling down the system, it undergoes a phase transition with a spontaneous symmetry breaking and an emergent total chirality that can be either 
positive or negative. 
A non-trivial band structure emerges for the electrons and when the Fermi energy lies in a gap, the systems becomes a Chern insulator 
whose Chern number depends on the sign of the spontaneous chirality. 
Note that this is achieved despite the fact that, as the magnetic system is bi-dimensional, it does not have a true long-range magnetic order. 
The second case
corresponds to a system which also shows a non-zero total chirality, 
but whose sign is governed by the orientation of the applied magnetic field.
As result, we can easily switch the Chern number of the filled band by just switching the orientation of the magnetic field. 
In both cases, we numerically investigate the density of states (DoS) and the value of the Hall conductivity, 
and in particular its evolution with the cooling temperature of the classical background. \\

\section{Hall conductivity in a chiral background}

\begin{figure}[htb]
\includegraphics[width=6cm]{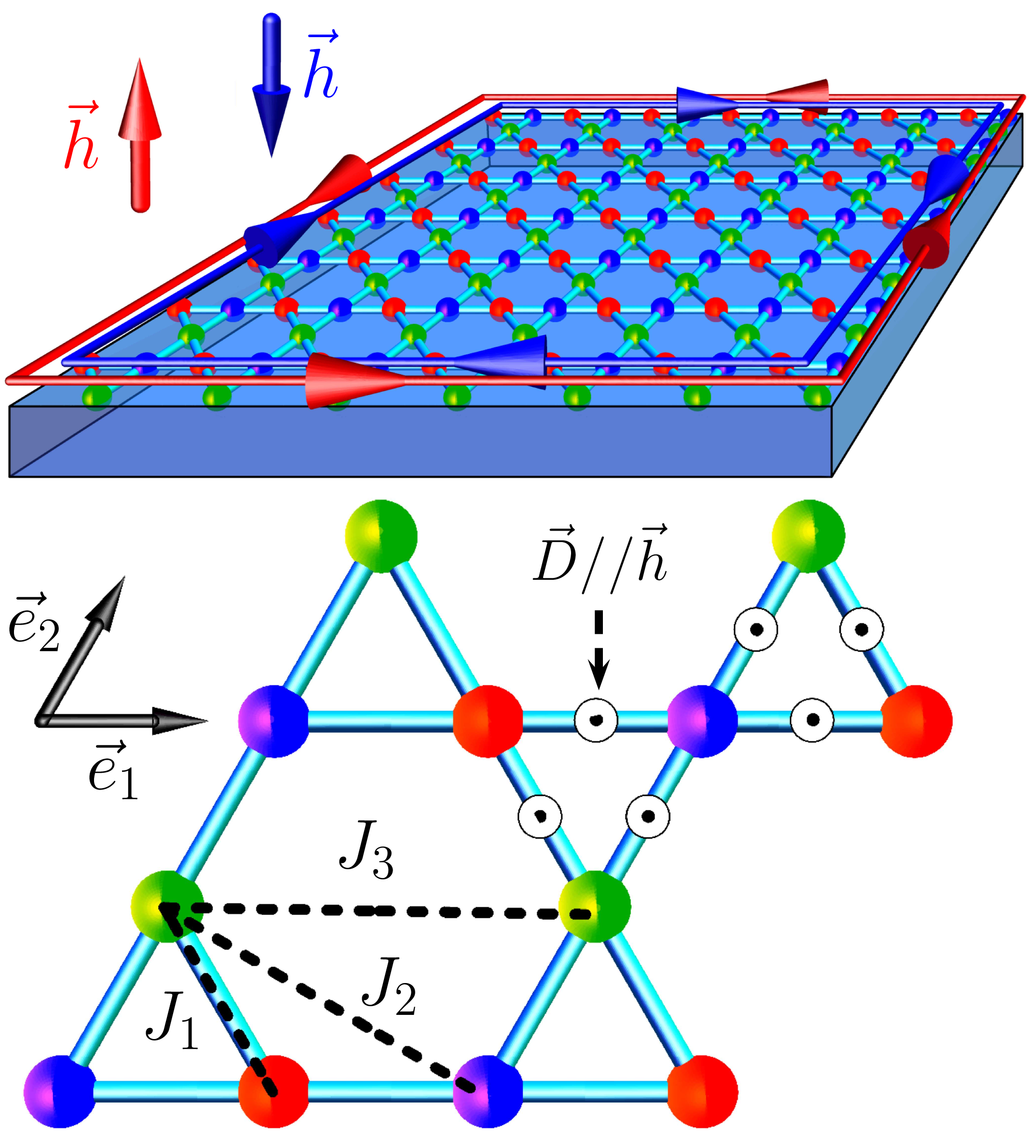}
\caption{(Top) Schematic representation of a topological transistor. 
The  spheres represent the sites of a (finite with edges) kagom\'e lattice
under an external magnetic field peperpendicular to the kagome layer. Different orientations of the magnetic 
field up (red) or down (blue) induce opposite mobility of the electrons. (Bottom) The kagom\'e lattice consisting of three sublattices
(red, blue and green spheres)  with Bravais vectors $\vec{e}_1=(1,0)$ and $\vec{e}_2=(\frac{1}{2},\frac{\sqrt{3}}{2})$. 
Dashed black lines   represent the first, second and third nearest neighbors exchange couplings, $J_1$, $J_2$ and  $J_3$ respectively.
$\vec{D}$  represents the Dzyaloshinskii-Moriya interaction parallel to the external magnetic field.} 
\label{fig:lattice}
\end{figure}

We consider a Kondo lattice model on the kagome lattice where the itinerant electrons are coupled with the classical spins by a 
Hund's coupling as,

\bea
H&=&-t\sum_{\langle \rv,\rv'\rangle,\sigma}(c^{\dagger}_{\rv,\sigma}c_{\rv',\sigma}+H.c.)-J_H\sum_{\rv}\Sp_{\rv}\cdot\Tp_{\rv}\nonumber\\
&&+H_S
\label{eq:Hamiltonian}
\eea
where $\Tp_{\rv}=\frac{1}{2}c^{\dagger}_{\rv,\mu}\vec{\sigma}_{\mu\nu}c_{\rv,\nu}$. Here, the first term is the hopping of itinerant electrons, 
where $c_{\rv,\sigma}$ ($c^{\dagger}_{\rv,\sigma}$) is the annihilation (creation) operator of an itinerant electron with spin $\sigma$ at $\rv$th site. 
The sum $\langle \rv, \rv'\rangle$ is taken over nearest neighbors (NN) sites
on the kagome lattice (see Fig. \ref{fig:lattice}), and $t$ is the NN hopping term. The second term is the onsite interaction
between localized spins and itinerant electrons, where $\Sp_{\rv}$ and $\Tp_{\rv}$ represent the localised spin and
itinerant electron spin at $\rv$-th site, respectively ($|\Sp_{\rv} | = 1$), and $J_H$ is the coupling constant. 
The last term corresponds to the pure magnetic Hamiltonian. 

We are interested in the limit of strong enough Hund coupling $J_H\gg t$. In this limit, although the spin of the hopping electron $\Tp_{\rv}$ is not conserved,
the electronic spectrum  clearly splits into a low energy and high energy band set 
where the spin of the electrons are roughly aligned parallel and antiparallel, respectively, to the local moment $\Sp_\rv$. 
If the magnetic background has non-zero chirality, this spin-orbit coupling results in an effective flux acting on the low energy band of the electronic 
sector \cite{Nagaosa_1,Taillefumier,Nagaosa_2}. Therefore, throughout this work we take $J_H/t=8$.

We investigate the transport properties of the model in Eq.  (\ref{eq:Hamiltonian}) by combining Monte-Carlo simulations for the 
classical magnetic sector and exact diagonalization for the quantum  electron gas.
Here, we obtain the classical magnetic background  using the  standard Metropolis algorithm combined with
overrelaxation (microcanonical) updates by studying a pure spin  model on a kagome lattice. Periodic boundary conditions
were implemented for a system of $N =3\times L^2$ sites ($L=12-36$). At  every magnetic field or temperature we discarded $1 \times 10^5$ hybrid 
Monte Carlo steps (MCS) for initial relaxation and spin configurations were collected during subsequent $2 \times 10^5$ MCS.
The different phases can be identified from the observation of real-space spin observables. In this case, we calculated  the thermal average of
scalar chirality, which is defined as the mixed product of three spins on a triangular  plaquette, $\chi_{\triangle}=\Sp_{\rv}\cdot(\Sp_{\rv'}\times\Sp_{\rv''})$.

In all this work all the  computations for the itinerant electrons for a  given classical spin configuration  are done at zero temperature, which 
corresponds to a vast separation between the electronic (hopping) energy scales and the Heisenberg energy 
scale of the classical system. 
We discuss the effect of thermal disorder in the magnetic background and 
we calculate the conductivity by means of the Kubo formula  which, at $T=0$,  is reduced to

\begin{eqnarray}
\sigma_{xy}(\varepsilon_F)
&=&
\frac{e^2}{\hbar L^2}\sum_{\epsilon_\mu < \varepsilon_F}\sum_{\epsilon_\nu \ge \varepsilon_F}
\frac{2\text{Im}\left([v_x]^{\mu\nu}[v_y]^{\nu\mu}\right)}{(\epsilon_\nu-\epsilon_\mu)^2}
\label{eq:kuboformulaW0}
\end{eqnarray}
where $\epsilon_F$ is the Fermi energy, $\epsilon_\mu$ is the eigenvalue of electronic Hamiltonian with the eigenvector $|\mu\rangle$. $[v_a]^{\mu\nu}=\langle \mu|\hat{v}_a|\nu\rangle$ ($a=x,y$) is the matrix elements of
the  velocity operator $\Vop=\frac{i}{\hbar}\left[\hat{H},\Rop\right]$, with the position operator being $\Rop=\sum_{\rv,\sigma}\rv\,c^{\dagger}_{\rv,\sigma}c_{\rv,\sigma}$. When the  Fermi energy $\epsilon_F$ lies inside the
gap, the Hall conductance is quantized as $\sigma_{xy}=e^2/h\sum_nC_n$ where the integers $C_n$ \cite{ChernNumbers} are the so-called Chern numbers connected with the extended (conducting) states.

\section{Magnetic XXZ model and spontaneous Chern insulator}

Our first model consists in an anisotropic XXZ Hamiltonian with first and second nearest neighbours 
exchange interactions:
\bea
H_S&=&\sum_{a=1}^2\sum_{\langle \rv,\rv'\rangle_a}J_a(\Sp^{\bot}_{\rv}\cdot \Sp^{\bot}_{\rv'}+\Delta_a S^z_{\rv}S^z_{\rv'})-h\sum_{\rv}S^z_{\rv}\quad
\label{eq:Hxxz}
\eea
where $\langle \rv, \rv'\rangle_a$ represents the first ($a=1$) and second ($a=2$) neighbours, with anisotropy $\Delta_a<1$ in the $z$ direction,
and $\Sp^{\bot}$ corresponds to the perpendicular ($xy$) component of the spin at  site $\rv$.

\begin{figure}[tbh]
\includegraphics[width=7.5cm]{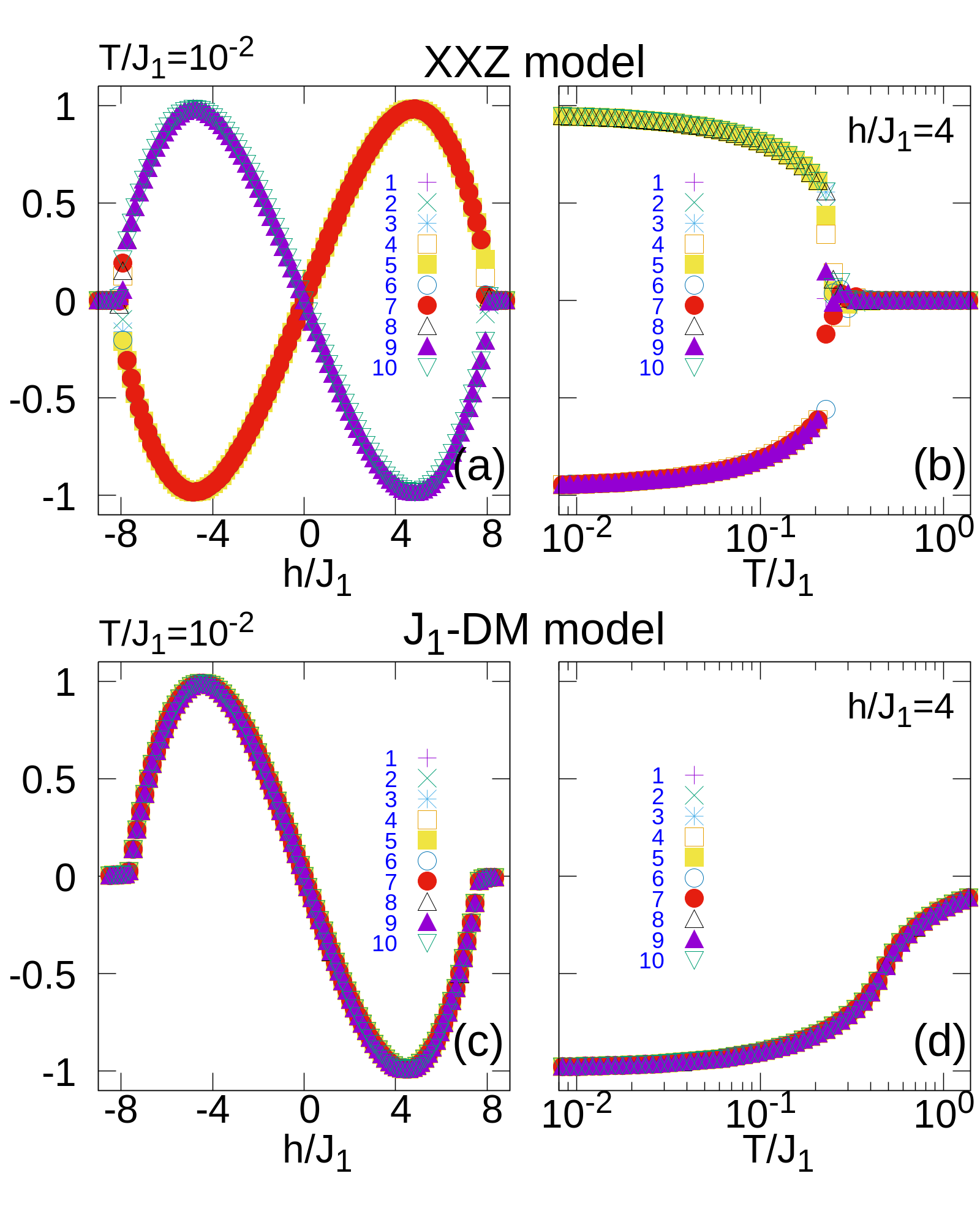}
\caption{Comparison of the normalised chirality per plaquette in 10 realizations of the MC simulations for the two models studied, (top) XXZ with $J_2/J_1=1/2,\Delta_1=\Delta_2=0.9$  and (bottom) $J_1-DM$ with $D/J_1=1/2$. 
Chirality as a function of the magnetic field at $T/J_1=10^{-2}$ (left) and as a function of temperature for $h/J_1=4$ (right)  
 We observe that while the XXZ model presents spontaneous 
chirality which changes in different realizations (a)-(b), in the $J_1-DM$ model the total chirality if fixed  by the direction of the magnetic field (c)-(d). } 
\label{fig:chirality}
\end{figure}

At low temperatures the system undergoes a phase transition in which the reflexion symmetry is spontaneously broken. 
The local magnetic order consists in a $q=0$ texture with non-zero chirality on each triangle of the kagome lattice.
This is precisely the configuration considered in Ref. [\onlinecite{Nagaosa_1}] for the first realisation of a Chern insulator due to a strong Hund's coupling.

\begin{figure*}
\includegraphics[width=17cm]{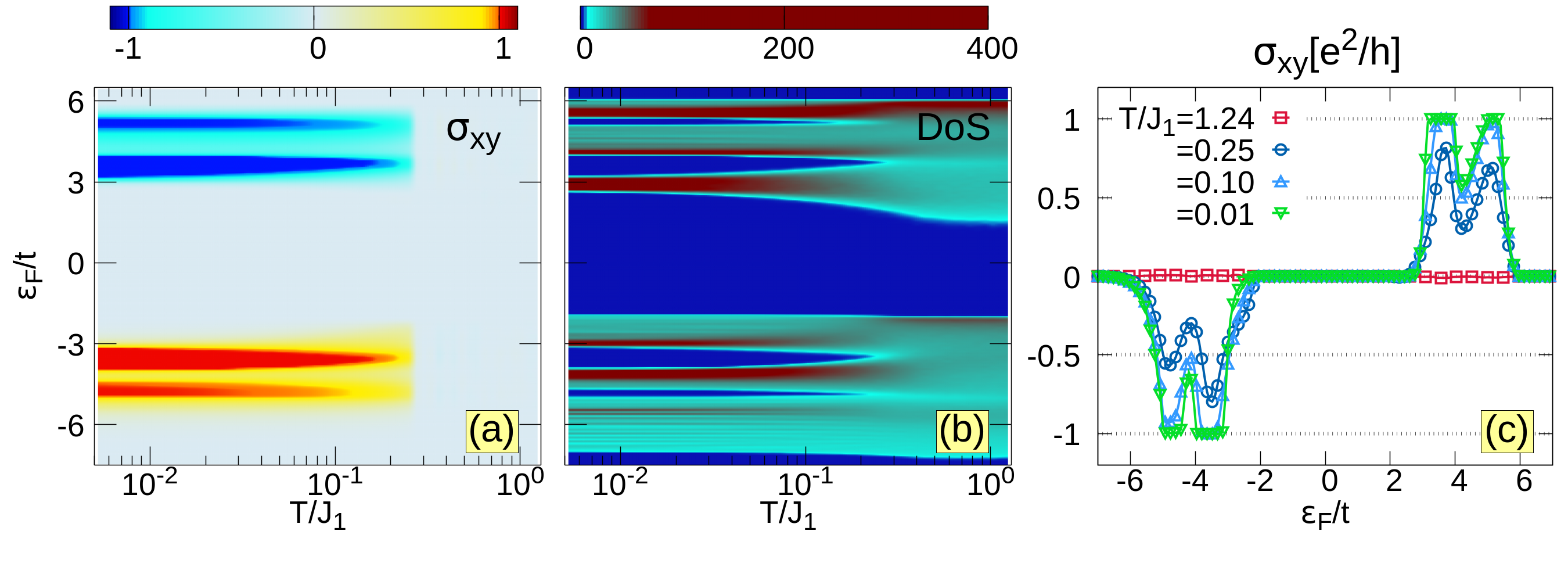}
\caption{XXZ model ($J_2/J_1=1/2,\Delta_1=\Delta_2=0.9$), at $h/J_1=4$. Density plot of (a) Hall conductivity (absolute value - in units of $e^2/h$) and (b) DoS vs $T/J_1$ ($x-$axis) 
and Fermi energy ($y$-axis). Results shown are the average of 100 realisations for $N=3888$. 
(c) Hall conductivity as a function of the Fermi energy for four different temperatures. } 
\label{fig:DoE}
\end{figure*}
\begin{figure}[tbh]
\includegraphics[width=7.5cm]{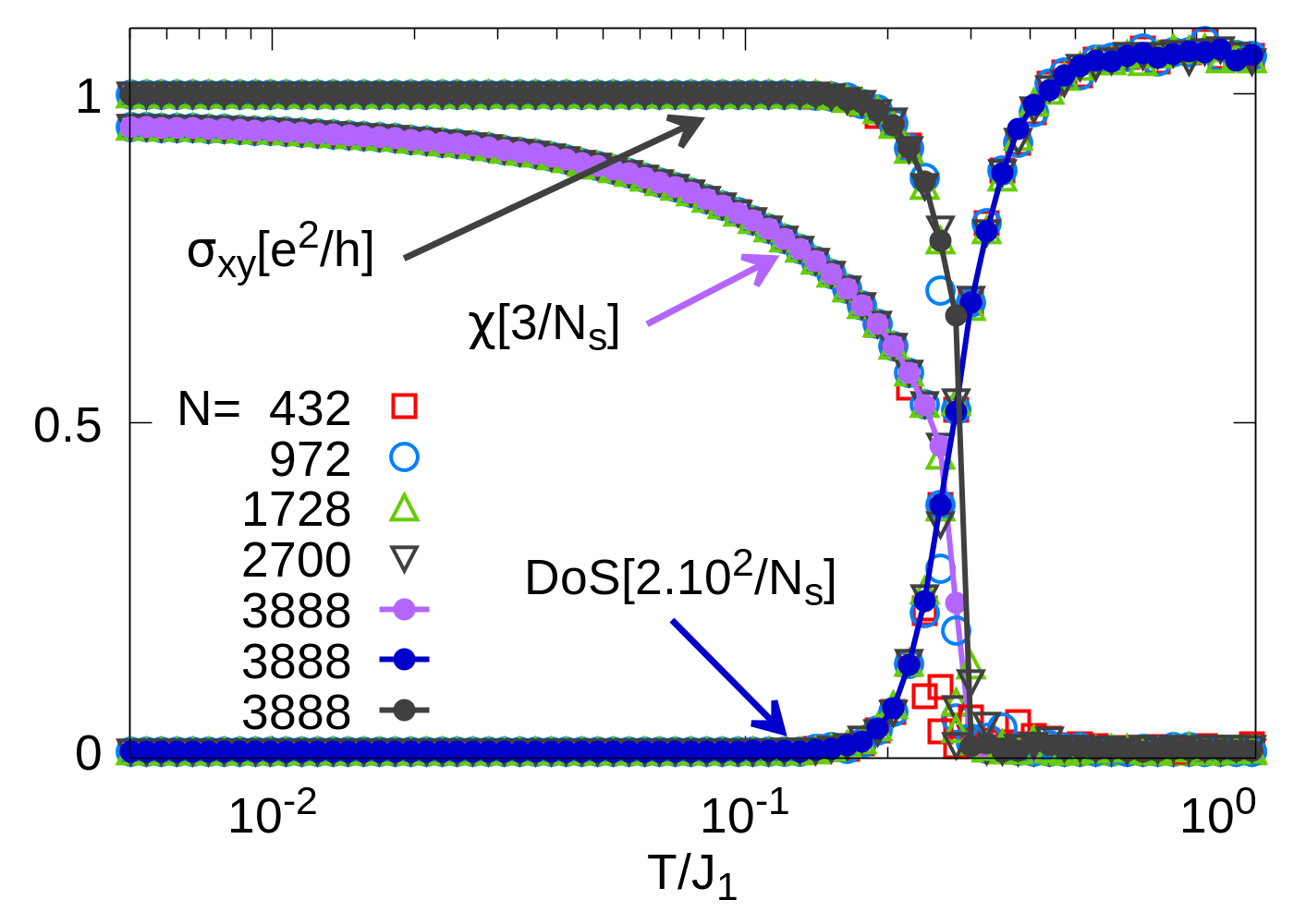}
\caption{ Hall conductivity (absolute value),
DoS and scalar chirality (absolute value) of the magnetic background as a function of temperature, for several system sizes for the 
XXZ model ($J_2/J_1=1/2,\Delta_1=\Delta_2=0.9$), at $h/J_1=4$, averaged over 100 realisations.} 
\label{fig:sizes_chi-dos-sxy}
\end{figure}
In  Figs. \ref{fig:chirality} (a) and (b)  we show the total chirality (per plaquette) $\chi=\frac{1}{L^2}\sum_{\triangle}\chi_{\triangle}$ as a function
of the magnetic field and temperature, respectively.
The appearance of a spontaneous chirality is not necessary related to the emergence of quasi-long-range-order of the magnetic structure which may appear,
via a standard Berezinsky-Kosterlitz-Thouless transition, at a lower temperature. 
Of course, the higher the temperature at which the classical spin system is thermalised, the higher is the disorder felt by the hopping electrons, 
and the more complicated the resulting spectral structure. This can be clearly seen in Fig.  \ref{fig:DoE} (b) where the DoS of the electronic system is plotted. 
However, at low temperatures ($T/J_1 \lesssim 10^{-1}$) a clear band and gap structure can be identified within the two decoupled Hund sectors.
Fig.  \ref{fig:DoE} (a) shows the value of the Hall conductivity which clearly takes quantised values when the Fermi energy lies within a gap.

A closer look at the transition reveals a very interesting scenario, as shown in Fig.\ref{fig:sizes_chi-dos-sxy}.
The emergence of the gap in the electronic sector, and of a non-zero and quantised Hall conductance,
seems to immediately appear with the onset of a total chirality of the background.
This seems to indicate that, despite the presence of disorder felt by the electrons because of the thermal fluctuations of the background, 
the  Chern insulator  emerges as soon as its ``driving force'', the chirality, gets a non-zero value. 
This makes the situation relatively different to what has been studied in the context of topological insulators in the presence of quenched disorder \cite{disorder+Chern}. 
In particular, the transition from a topological to a non-topological state may be of a quite different nature. To study the system size dependence in these transitions, in Fig.\ref{fig:sizes_chi-dos-sxy} we plot the absolute value of the chirality, the Hall conductance and the DoS for several system sizes. It can be clearly seen that the behaviour is robust against finite size effects. 

\section{Field induced topological transistor}
\begin{figure}[htb]
\includegraphics[width=8.5cm]{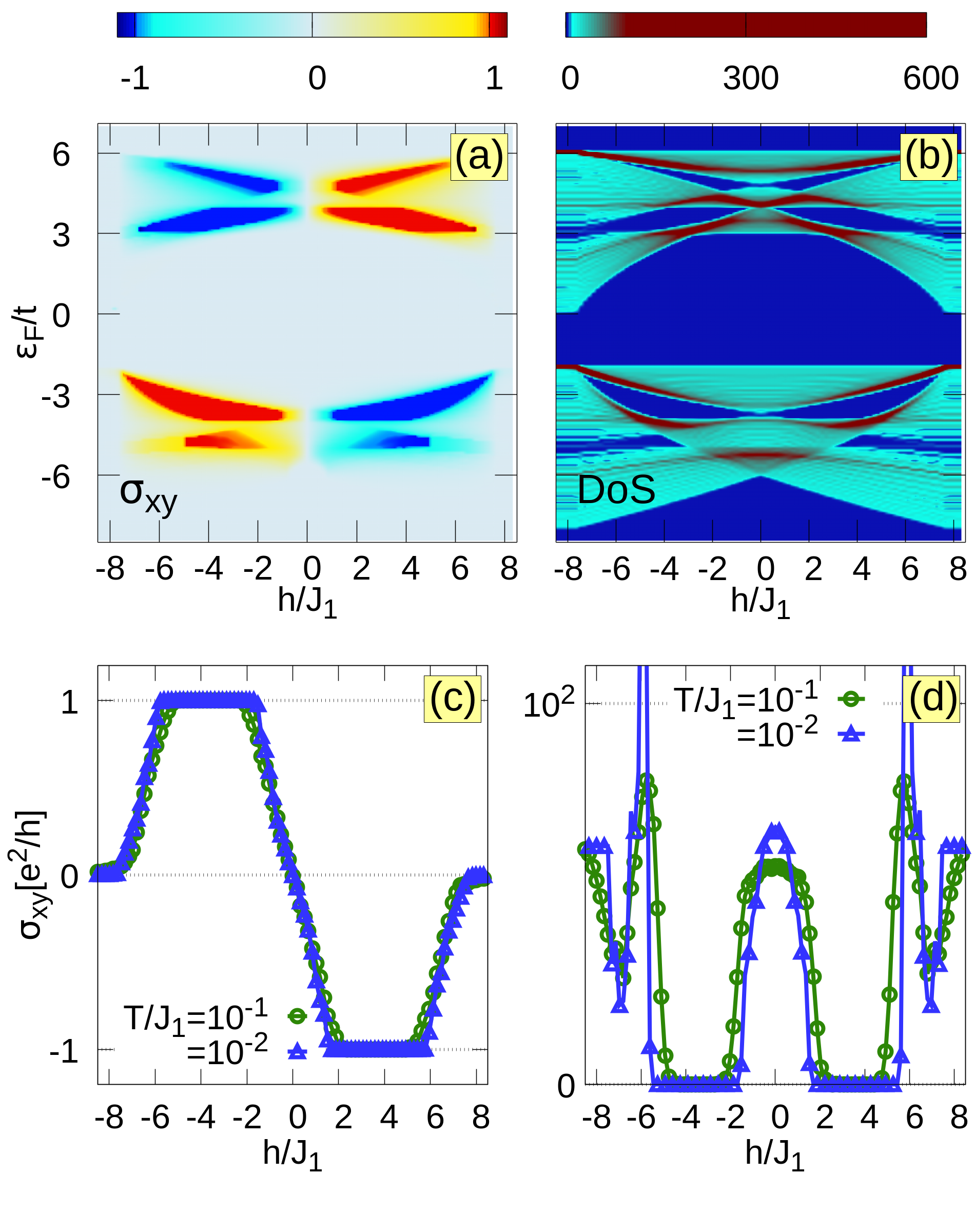}
\caption{$J_1-DM$ model ($D/J_1=1/2$). Density plot of (a) Hall conductivity (in units of $e^2/h$) and (b) DoS vs external magnetic field ($x-$axis) and Fermi energy ($y-$axis)
for $T/J_1=10^{-2}$.  Results shown are the average of 100 realisations for $N=3888$. Hall conductivity (c) and DoS (d) vs external magnetic field (at $\varepsilon_F/t=-3.5$) for two temperatures.
We observe the integer values for $\sigma_{xy}$ at the zero values of DoS} 
\label{fig:Cond_Hall}
\end{figure}

For our second model we consider for the classical spin sector a Heisenberg Hamiltonian in a magnetic field including  an out-of-plane Dzyaloshinskii-Moriya (DM)
interaction, parallel to the external magnetic field:
\bea
H_S&=&\sum_{\langle \rv,\rv'\rangle_1}J_1\Sp_{\rv}\cdot\Sp_{\rv'}+{\bf D}\cdot(\Sp_{\rv}\times\Sp_{\rv'})-h\sum_{\rv}S^z_{\rv}\quad
\label{eq:Hj1j2dm}
\eea
This system is known to give rise to a $q=0$ local magnetic order with a non zero chirality which depends on the sign of the magnetic field \cite{Elhajal},
as can be seen from Figs. \ref{fig:chirality} (c) and (d). For strong enough values of the magnetic field there is band structure with gaps emerging 
(besides the Hund's gap) as can be seen in Fig. \ref{fig:Cond_Hall} (b).
When the Fermi energy is chosen to lie within a gap, the Hall conductivity is clearly quantised, 
and can be "tuned" with the orientation of the magnetic field (Figs \ref{fig:Cond_Hall} (a) and (c)). It is important to stress that in an experimental context, the magnetic field applied to the magnetic background needed to obtain the desired effect is much 
lower than the one needed for realizing a genuine quantum Hall effect in the electronic sector. 

\section{Discussion and perspectives}
In this work we have studied the interplay between a classical frustrated magnetic system and electronic degrees of freedom which can become,
under some circumstances, a Chern insulator. The study of this kind of system dates back to almost twenty years ago \cite{Nagaosa_1}.
Here we have taken into account the thermal fluctuations of the magnetic background and we have investigated 
under which circumstances and range of temperatures one can recover the physics of the perfect background. 
Moreover, we have studied two models that show an interesting behaviour by tuning either the temperature or the magnetic field of the classical magnetic background.
In the first example, we showed that by lowering the temperature we can trigger a phase transition 
with a spontaneous symmetry breaking and the appearance of a non-zero total scalar chirality, 
triggering in turn the appearance of a interesting band structure for the coupled electronic system and the presence of a ``spontaneous'' Chern insulator. 
The second model we have studied also provides the realisation of a Chern insulator, 
but in this case the Chern number of the filled bands (and as so the direction of the chiral edge modes) can be switched by inverting the sign of the magnetic field. 

Our results open many perspectives in studying this kind of system. First, thermal fluctuations in the classical background play the role of disorder 
felt by the hopping electrons. As so, tuning the temperature amounts for tuning the degree of disorder in the electronic system, giving rise
to the appearance of localised states at the band edges and even closing up the gaps by further increasing the temperature. 
In fact, the evolution from a topological state to a (non topological) Anderson insulator, as well as the nature of the transitions involved, is a fundamental question for the physics of the quantum Hall effect and has been widely studied \cite{disorder+Hall}. These same questions arise in the study of topological insulators in the presence of impurities, a subject which is actually a very active field of research \cite{disorder+Chern}.
The situation in our context is relatively different to what has being studied before, as there is no quenched disorder in the system, but the disorder is induced by the temperature in the magnetic background. Moreover, there appears to a be a seemingly simultaneous transition to a chiral state for the spin background and to a Chern insulator for the electronic sector. A more detailed study of the nature of this transition is certainly a very interesting open issue. It would be also interesting to investigate if there are similar systems that would show two well separated transitions, a first one for the emergence of the chirality, and a second one, at a lower temperature, where the electronic sector undergoes a transition to a gapped topological state, which would be a more similar situation to the models studied in the literature \cite{disorder+Chern}. 
Still concerning disorder, it is possible to generate a classical chiral spin liquid \cite{Flavia+Pierre},
where even the local magnetic order is washed out, but where a total non-zero chirality is present.
This system may also present an interesting background for studying the electronic properties. 

Another important issue concerns the interplay between the electronic and magnetic degrees of freedom. 
In this work we have assumed a huge separation of energy scales between the two sectors, 
and as so we have computed all the electronic observable and zero temperature. 
The next step would of course consist in studying the case where the energy scales of the two sectors are closer. 
Not only it is possible to compute the electronic observables at non-zero temperature, 
but it is also possible to implement a feed-back mechanism of the electrons to the magnetic background. This is also a direction that promises a very rich and interesting phenomenology.

Furthermore, our study of the interplay between thermal disorder and QAH effect on the XXZ and $J_1-DM$ models opens the door for a deep analysis on others 2D antiferromagnetic 
lattices with noncoplanar commensurate
and incommensurate orders as in FeCrAs \cite{Wu2009} (where small spin-orbit effects play an important role) and the layered (itinerant)
kagome lattice Fe$_3$Sn$_2$\cite{Fenner2009}. Another possibility is the study of the effect of nonmagnetic impurities 
on frustrated Kondo-lattice models, since  the competition 
between thermal fluctuations and the site disorder in pure spin models can stabilize non coplanar states  \cite{Mike2013}.
At last, another very promising issue is to consider quantum fluctuations of the magnetic background, and how the topological nature of the electronic sector could facilitate the emergence of a topological spin liquid in the magnetic sector \cite{Makh_Puj, Hsieh}.
In all these cases, the perspective of an experimental setup for studying in a controlled way the transition to topological states is a very promising issue.     

\section*{Acknowledgments}
It is a big pleasure to acknowledge Revaz Ramazachvili for suggesting us to investigate the possibility of a topological transistor.
H.D.R. and F.A.G.A. thank the Laboratoire de Physique Th\'eorique
(LPT) in Toulouse for their hospitality.
H.D.R. and F.A.G.A. are partially supported by PIP 2015-0813 CONICET and SECyT-UNLP. 
H.D.R. acknowledges support from PICT 2016-4083. This research is funded in part by a
QuantEmX grant from ICAM and the Gordon and Betty Moore Foundation
through Grant GBMF5305 to H.D.R..

\end{document}